\documentclass{PoS}

\usepackage{epsfig}
\usepackage{multicol}
\usepackage{pstricks,pst-grad}
\usepackage{hepunits}
\usepackage{xcolor}
\usepackage{tikz}
\usepackage{rotating}
\usepackage{slashed}

\newcommand{\beq}{\begin{equation}}
\newcommand{\eeq}{\end{equation}}
\newcommand{\bfig}{\begin{figure}[ht]\begin{center}}
\newcommand{\efig}{\end{center}\end{figure}}

\newcommand{\msbar}{{\overline{\rm MS}}}

\newcommand{\rismom}{{\rm RI/SMOM}}

\newcommand{\rismomqslash}{{\rismom_{\slashed{q}}}}
\newcommand{\rismomgamma}{{\rismom_{\gamma_{\mu}}}}

\newcommand{\NDR}{\msbar{\rm [NDR]}}

\newcommand{\tree}{{\rm tree}}

\title{The Kaon Bag Parameter at Physical Mass}

\ShortTitle{The Kaon Bag Parameter at Physical Mass}

\author{\speaker{Julien~Frison}$^1$, Peter~Boyle$^1$, Norman~H.~Christ$^2$, Nicolas~Garron$^3$, Robert~Mawhinney$^2$, Chris~T.~Sachrajda$^4$, Hantao~Yin$^2$\\
        School of Physics \& Astronomy, University of Edinburgh, EH9 3JZ, UK$^1$, \\
	Physics Department, Columbia University, New York, NY 10027, USA$^2$, \\
	School of Mathematics, Trinity College, Dublin, Ireland$^3$, \\
        School of Physics and Astronomy, University of Southampton, SO17 1BJ, UK$^4$\\
        E-mail: \email{jfrison@ph.ed.ac.uk}, 
        \email{paboyle@ph.ed.ac.uk}, 
	\email{nhc@phys.columbia.edu}, 
	\email{ngarron@maths.tcd.ie}, 
	\email{rdm@physics.columbia.edu}, 
        \email{cts@soton.ac.uk}, 
	\email{yinnht@phys.columbia.edu}}

\author{RBC-UKQCD Collaboration}

\abstract{We present preliminary results for the calculation of the Kaon Bag parameter $B_K$ in $N_f=2+1$ lattice QCD, using 
M\"obius Domain Wall Fermion ensembles generated by the RBC-UKQCD collaboration. This computation is done directly at physical 
meson masses, so that we do not have to rely on chiral perturbation theory or any other mass extrapolation. In parallel, 
the four-quark operator is renormalised through the Rome-Southampton technique. Finally, we compare 
our value with previous results and draw some conclusions about the remaining dominant contributions in our error budget.}

\FullConference{31st International Symposium on Lattice Field Theory - LATTICE 2013\\
		July 29 - August 3, 2013\\
		Mainz, Germany}

\begin{document}

\section{Introduction}

The mixing of neutral mesons such as the Kaon plays an important role in $CP$-violation, as the so-called {\it box-diagrams} involve 
terms like $V_{xs}^*V_{xd}$. However, those perturbative diagrams cannot deal with the confinement of ingoing and outgoing quarks. Nevertheless, 
an Operator Product Expansion (OPE) allows a separation between the perturbative part and a non-perturbative QCD part that can be computed on
the lattice. Lattice computation have already been very successful on this path (see \cite{FLAG,LatticeAverages} for a review, and 
 \cite{PhysRevD.84.014503,PhysRevD.87.094514} for previous results of our collaboration), and what is
at stake is now an improvement of the uncertainties. 

In the context of lattice computation, we traditionally express the $K^0-\bar K^0$ matrix element as a ratio normalised by the
result of Vacuum Saturation Approximation. 
On a practical side we therefore have to compute three-point and two-point functions, and fit a ratio on ranges that form a plateau converging to the
asymptotic constant. 

Then, this result has to be brought to a renormalisation scheme which is common with the perturbative computations. This has been done
through the Rome-Southampton method and using several intermediate schemes to evaluate the systematic error coming from this part of 
the computation. In order to improve the convergence to the perturbative regime, non-exceptional momenta have been used and the matching
is done at a scale of $3\ \GeV$.

The present calculation, which is preliminary, has been done on the new $N_f=2+1$ M\"obius Domain-Wall Fermion (MDWF) ensembles of the RBC-UKQCD collaboration, generated directly at
the physical pion mass. 
We used two ensembles with roughly the same physical volume ($M_\pi L\simeq 3.8$) but different lattice spacing: a $48^3\times 96$ ensemble at $\beta=2.13$
and a $64^3\times 128$ ensemble at $\beta=2.25$. Finite-volume errors are estimated from older ensembles and are not a crucial issue.
The bare $B_K$ ratios have been computed with respectively $60$ and $21$ configurations of those ensembles, while $11$ and $6$ configurations were
sufficient for the Non-Perturbative Renormalisation (NPR) of these preliminary results. All of these measurements will ultimately be performed on larger data sets as part of a forthcoming publication, and as such this work is prelimary. 

\section{Bare $B_K$ ratio}

For the bare ratio, all contractions have been computed using Coulomb gauge fixed wall sources and the All Mode Averaging (AMA) method. The
latest combines propagators computed with different precisions (here $10^{-4}$ and $10^{-8}$) in order to improve signals at a minimal cost \cite{Blum:2012uh}.

The ratio
\begin{equation}
B_K^{bare}(t,\Delta t) = \frac{\langle K^0(\Delta t)\mid {\cal O}_{VV+AA}^{\Delta S=2}(t)\mid \bar K^0(0)\rangle_{AMA}}{\frac{8}{3}\langle K^0(\Delta t-t)\mid A_0(0)\rangle_{AMA}\langle A_0(t)\mid \bar K^0(0)\rangle_{AMA}} 
\end{equation}
is fitted to a constant 
on the two-dimensional space $(t,\Delta t)$, 
The fit domain is chosen as $\Delta t\in [\Delta t_{min},\Delta t_{max}]$ 
and
$t\in [t_{margin},\Delta t-t_{margin}]$ 
(plus symmetric times). 

For the $48^3$ ensemble we choose the domain $(t_{margin},\Delta t_{min},\Delta t_{max})=(8,20,32)$, giving
\beq
B_K^{48} = 0.5840(9)
\eeq
for an uncorrelated fit
with $\chi^2/dof = 0.76$, while still giving an acceptable correlated $\chi^2/dof = 1.68$. 
On the $64^3$ ensemble stabilising the correlated fit becomes impossible and we perform an uncorrelated fit on the domain $(t_{margin},\Delta t_{min},\Delta t_{max})=(10,25,40)$,
giving (with $\chi^2/dof = 0.64$, see Fig.~\ref{fig:bk64_fit})
\beq
B_K^{64} = 0.5627(7) .
\eeq

\bfig
  \includegraphics[width=0.7\linewidth]{bk64_fit}
  \caption{We show a part of the fit on the $64^3$ ensemble. Green lines represent the result of the fit with its error bars. The fit is actually two-dimensional and
           we only represent one $\Delta t$ timeslice for readibility. }
  \label{fig:bk64_fit}
\efig

\section{Non Perturbative Renormalisation}

\begin{figure}[ht]
\begin{center}
  \hfill
  \includegraphics[width=0.3\linewidth]{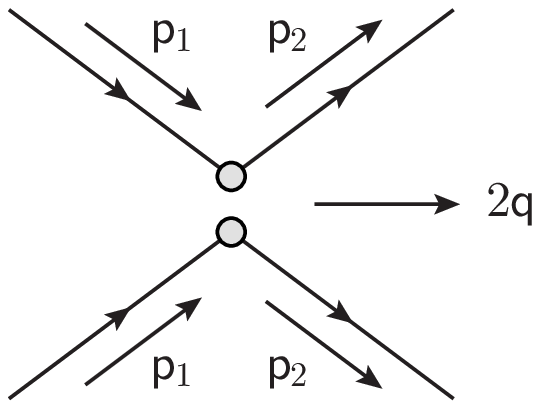}
  \hfill
  \includegraphics[width=0.3\linewidth]{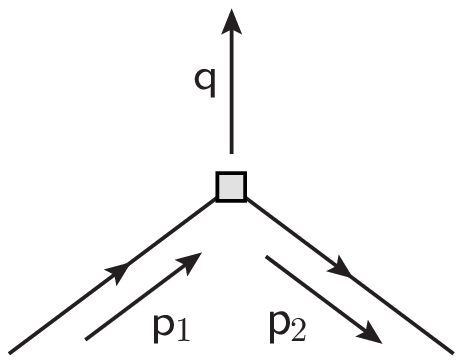}
  \hfill
  \hfill
  \caption{Four-quark Green's functions $G_{VV+AA}$ with momenta defining the $\rismom$ non-exceptional scheme (left), and the scheme definition for the bilinears (right). 
    The momenta are chosen such that $p_1^2=p_2^2=q^2$, where $q=p_1-p_2$. 
    An additional momentum of $2q$ (resp. $q$) leaves the operator as indicated by the arrow. Every momentum or subset of momenta is of order $q$ (never zero), which improves
    the pertubative running. }
  \label{fig:cctree}
\end{center}
\end{figure}

We have to renormalise a four-quark operator, which hopefully renormalise multiplicatively in a Domain-Wall discretisation. In the RI/SMOM scheme this is done
through the computation of four-quark Green's functions such as Fig.~\ref{fig:cctree}, where ingoing and outgoing quarks use Landau gauge-fixed wall sources 
with momenta on those walls. 

We note $\Gamma_{VV+AA}$ the associated amputated Green's function. The scheme is completely defined once we choose a projector $P$ allowing us to write the
renormalisation condition
\beq
P\Gamma_{VV+AA}({\cal O}^{RI}_{VV+AA}) = P\Gamma_{VV+AA}({\cal O}_{VV+AA})\mid_\tree .
\eeq
We will use the two choices
\begin{eqnarray}
P^{\gamma}_\mu &=&  \delta_{ij}\delta_{kl}\left[(\gamma_\mu)_{\beta\alpha}(\gamma_\mu)_{\delta\gamma}+(\gamma_\mu\gamma_5)_{\beta\alpha}(\gamma_\mu\gamma_5)_{\delta\gamma}\right]\\
P^{\slashed{q}} &=& \delta_{ij}\delta_{kl}\left[(\slashed{q})_{\beta\alpha}(\slashed{q})_{\delta\gamma}+(\slashed{q}\gamma_5)_{\beta\alpha}(\slashed{q}\gamma_5)_{\delta\gamma}\right]/q^2
\end{eqnarray}
defining respectively the schemes $\rismomgamma$ and $\rismomqslash$, where $i,j,k,l$ contract with the colour indices of $\Gamma_{VV+AA}$ and $\alpha,\beta,\gamma,\delta$
with its spin indices. 

This condition gives us the factor
\beq
\frac{Z_{VV+AA}}{Z_q^2} = \frac{ P\Gamma_{VV+AA}({\cal O}_{VV+AA})\mid_\tree }{ P\Gamma_{VV+AA}({\cal O}^{lat}_{VV+AA}) }
\eeq
and we do the same for bilinear operators to get
\beq
\frac{Z_{A}}{Z_q} = \frac{ P\Gamma_{A}({\cal O}_{A})\mid_\tree }{ P\Gamma_{A}({\cal O}^{lat}_{A}) }
\eeq
so that $Z_q$ cancels in the ratio
\beq
Z_{B_K} = \frac{Z_{VV+AA}}{Z_A^2} .
\eeq

\begin{figure}[ht]
\begin{center}
  \includegraphics[width=0.7\linewidth]{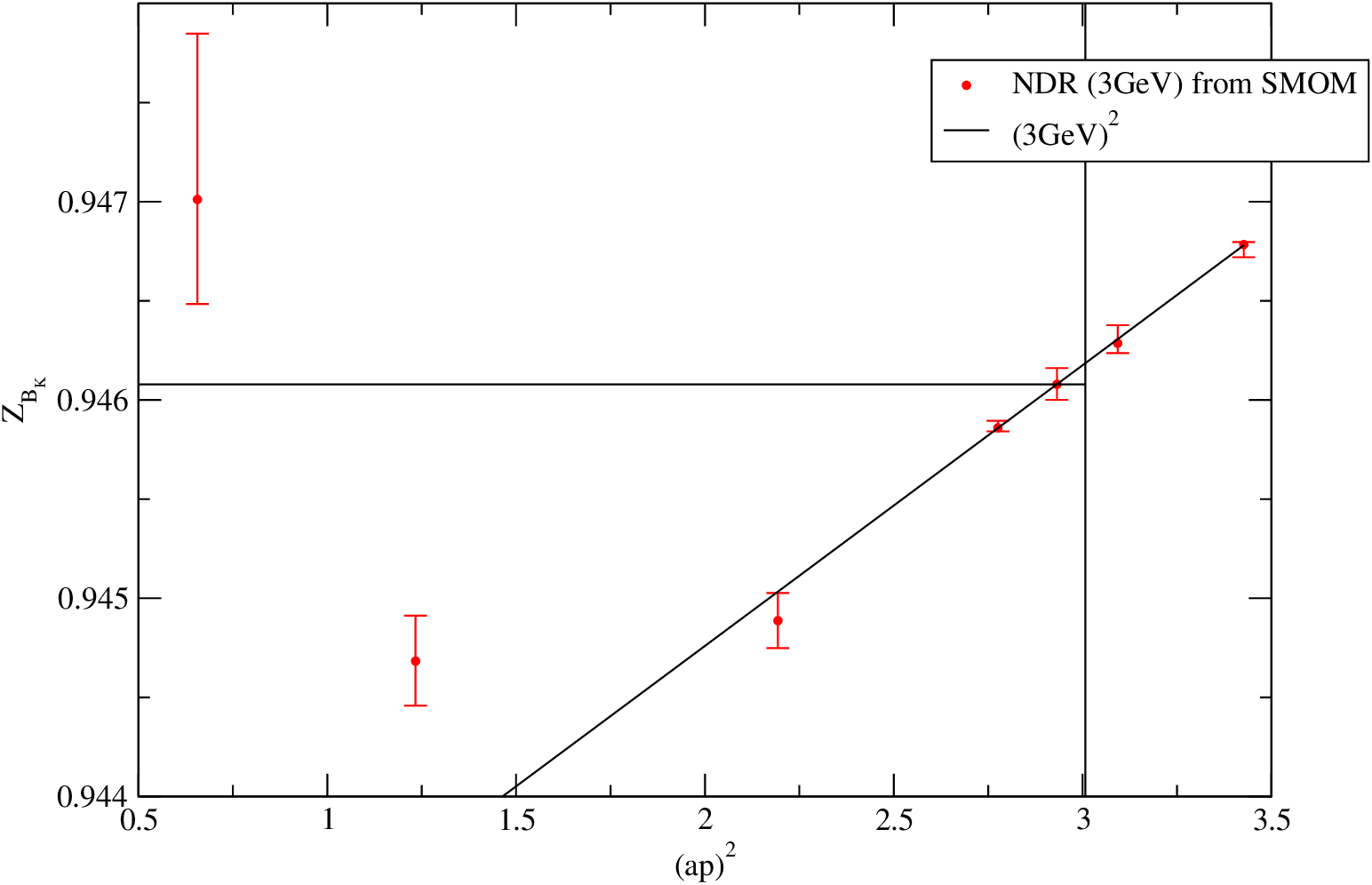}
  \caption{Renormalisation factor $Z_{B_K}$ in $\NDR(3\ \GeV)$ from intermediate scheme $\rismomqslash$, with $48^3$ ensemble. The vertical line 
           shows the $(ap)^2$ value corresponding to
           the scale $|p|=3\ \GeV$, while the horizontal line show the value for the closest simulated point relative to this scale. The slope of the linear fit
           is one of the ways to evaluate the systematic error due to not having an infinitely large Rome-Southampton window, although we prefer using the difference
           between different intermediate schemes at the end of the computation. 
Only the two points close to $3\ \GeV$ are actually used, the other momenta having very low statistics. 
}
  \label{fig:extrap_psq}
\end{center}
\end{figure}

Those renormalisation factors, computed on 
the momenta $[p,0,p,0]$ with $p=9.25$ and $p=9.5$ thanks to boundary twisting,
are run and matched to the pertubatively-defined $\NDR$ at scale $3\ \GeV$, as shown
in Fig.~\ref{fig:extrap_psq} with other results summarized in Tab.~\ref{tab:Zresults}. 


\begin{table}
\begin{center}
\begin{tabular}{|c|c|c|c|}
\hline
$\beta$ & scheme & This work ($48^3$ and $64^3$) & Previous results ($24^3$ and $32^3$)\\ 
\hline
$2.13$ & $\NDR(\gamma,3GeV)$ & $ 0.91764(4)$ & $0.91983(10)(51)(180)(3)$ \\
$2.13$ & $\NDR(\slashed{q},3GeV)$ & $0.94611(3)$ & $0.94672(11)(84)(63)(12)$\\
$2.25$ & $\NDR(\gamma,3GeV)$ & $0.94346(5)$ & $0.94284(17)(49)(48)(2)$ \\
$2.25$ & $\NDR(\slashed{q},3GeV)$ & $0.96722(2) $ & $0.96698(13)(86)(9)(2)$ \\
\hline
\end{tabular}
\end{center}
\caption{$Z_{B_K}$ results from different ensembles and schemes. On the right column we remind previous results with a very similar action\cite{PhysRevD.84.014503},
where here the first number is statistical error, then scale setting error, then $m_s$ errors, then $Z_A-Z_V$ errors. Those number agree within systematic
errors, and the statistical error has been drastically reduced even with a small number of configurations. }
\label{tab:Zresults}
\end{table}
%

\section{Preliminary Continuum Extrapolation}

After multiplying our bare $B_K$ by those values for the two schemes $(\gamma,\gamma)$ and $(\not q, \not q)$, we can extrapolate them to
the continuum, removing ${\cal O}(a^2)$ effects : 
\begin{eqnarray}
B_K^{\gamma,\gamma} &=& 0.5241(11)_{stat48}(17)_{stat64}(4)_{a48}(7)_{a64}\\
B_K^{\not q, \not q} &=& 0.5332(12)_{stat48}(17)_{stat64}(7)_{a48}(12)_{a64},
\end{eqnarray}
where we have separated errors coming from first the $B_K$ value on the $48^3$ lattice, then on the $64^2$ lattice, and then the errors propagated from each scale setting. 
The difference between those two values is an estimate of the systematic error due to the perturbation conversion from intermediate MOM schemes to $\msbar$. Concerning the
central value any combination of these figures could theoretically be justified, however we choose to use the $(\not q,\not q)$ scheme as
the reference as we observe a better perturbative matching from it. 
Including the estimation of finite-volume errors from our earlier
ensembles\footnote{This is a conservative estimation, since the application of ChPT\cite{Becirevic:2003wk} suggests that this error is now much reduced. }, we then 
quote the following \emph{preliminary} result: 

\begin{equation}
B_K^{\overline{\mathrm{MS}}}(3\ \GeV) = 0.533(3)_{stat}(0)_{\chi}(3)_{FV}(11)_{NPR}
\end{equation} 

\begin{figure}[ht]
\vspace*{-0.75cm} 
\begin{center}
  \includegraphics[width=0.7\linewidth]{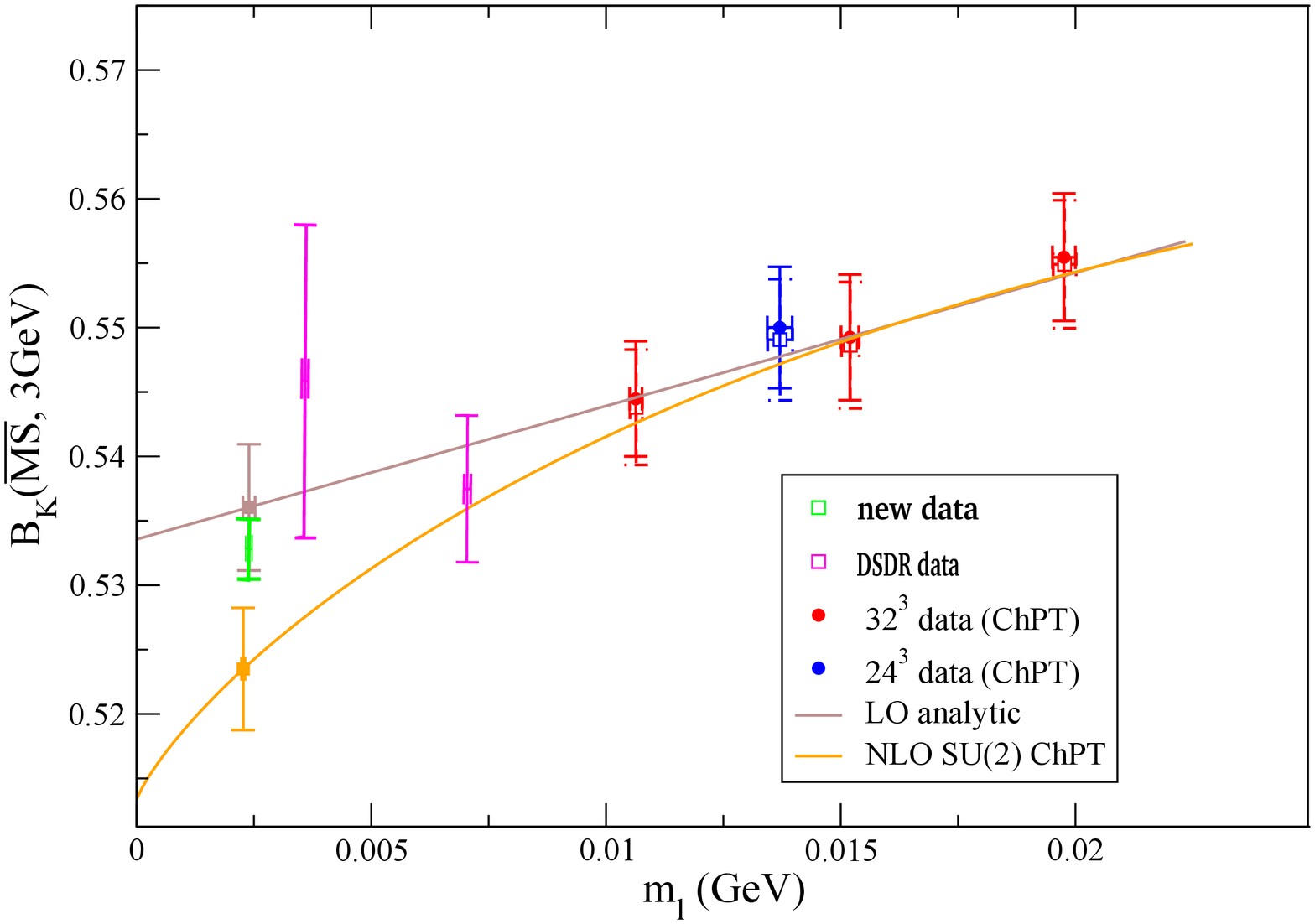}
\vspace*{-0.75cm} 
  \caption{We present the results of this study, in green, compared to previous results used in a chiral extrapolation. The fits we show are only based on
blue and red points, so that the systematic error separating the two formulae is very visible, although the purple data already suppressed much of this 
problem\cite{PhysRevD.87.094514}. In the present study, on top of improving the statistical error, those systematics are completely removed. }
  \label{fig:chiral_plot}
\end{center}
\end{figure}

\section{Conclusion}

This value has to be compared to the previous results of our collaboration in \cite{PhysRevD.87.094514}
\begin{equation}
B_K^{\overline{\mathrm{MS}}}(3\ \GeV) = 0.535(8)_{stat}(7)_{\chi}(3)_{FV}(11)_{NPR}
\end{equation}
and the older \cite{PhysRevD.84.014503}
\begin{equation}
B_K^{\overline{\mathrm{MS}}}(3\ \GeV) = 0.529(5)_{stat}(15)_{\chi}(2)_{FV}(11)_{NPR} .
\end{equation}
Although those result are perfectly compatible within error bars, we saw a slight increase of the central value with the inclusion of the near-physical mass ensembles, 
which reduced the gap between Taylor and ChPT fits and discarded large chiral logarithmic effects. The current analysis clearly confirms this phenomenum 
seen in \cite{PhysRevD.87.094514}.

We also provide a value in the Renormalisation Group Independent (RGI) scheme in Tab.~\ref{tab:RGI} in order to allow for a comparison with results of some other 
collaborations.

\begin{table}
\begin{center}
\begin{tabular}{|c|c|}
\hline
Collaboration & $\hat B_K$\\
\hline
\hline
This work (preliminary) & $0.755(4)(15)$\\
\hline
RBC-UKQCD'12 & $0.758(11)(19)$\\
RBC-UKQCD'10 & $0.749(7)(26)$\\
BMW'11 & $0.773(8)(9)$\\
SWME'11 & $0.716(10)(35)$\\
Aubin'09 & $0.724(8)(29)$\\
ETM'10 ($N_f=2$) & $0.729(25)(17)$\\
\hline
\end{tabular}
\end{center}
\caption{${\rm RGI}$ value for $B_K$ derived from different previous works. Other values might be found in \cite{FLAG,LatticeAverages}
.}
\label{tab:RGI}
\end{table}

We note that most determinations have a central slightly lower than the value we present. We suspect this could be related to ChPT fits outside their range of validity, and also the use of only $Z$ factors from schemes whose running is 
less well described by perturbation theory than $\rismomqslash$. 

Most of the works also are, like this one, dominated by systematic errors. In our case, now that data down to the physical pion mass has eliminated the
main systematic error, mass extrapolation, nearly all of the error is due to NPR. 

Therefore we have shown that our new data and techniques provide an important improvement of the $B_K$ value, leading 
to $B_K^{\overline{\mathrm{MS}}}(3\ \GeV) = 0.533(3)(11)$. This is still preliminary, and in particular would need more statistics on NPR to be confident 
of the currently negligible statistical error. The 
direct simulation at physical pion mass has been a very important
ingredient, eliminating the dominant systematic error. The very obvious conclusion is now that what is important is improving the
non-perturbative renormalisation. This would in particular imply determining $Z_{B_K}$ at higher energy scales
, which in turn would imply simulating on finer and finer lattices. 
Renormalisation on finer lattices could be achieved on small volumes with step scaling. This includes running over a charm threshold since
it is important to match to perturbation theory at scales above those where the real world charm becomes relevant.
This will be a major objective for the next simulations. Additionally, the determination of $Z$ factors can also be improved by
designing schemes with a better matching, a direction that has already been intensively studied by our collaboration, with improvements of
the RI-MOM scheme already used in this work.

\section*{Acknowledgements}

The authors gratefully acknowledge computing time granted through the STFC funded DiRAC facility (grants ST/K005790/1, ST/K005804/1, ST/K000411/1, ST/H008845/1). PAB acknowledges support from STFC Grant ST/J000329/1 and was also supported by the European Union under the Grant Agreement number 238353 (ITN STRONGnet).
Critical to this calculation were the Blue Gene/Q 
computers at the Argonne Leadership Computing Facility
(DOE contract DE-AC02-06CH11357) as well as the DOE 
USQCD and RIKEN BNL Research Center Blue Gene/Q computers 
at the Brookhaven National Lab.  NHC was supported in
part by US DOE grant DE-FG02-92ER40699.

%
\bibliographystyle{hunsrt_etal}

\bibliography{proceedings.bib}

\begin{thebibliography}{1}

\bibitem{FLAG}
G.~Colangelo, S.~D\"urr, A.~J\"uttner, et~al.
\newblock Review of lattice results concerning low-energy particle physics.
\newblock {\em The European Physical Journal C}, 71(7):1--76, 2011.

\bibitem{LatticeAverages}
Jack Laiho, E.~Lunghi, and Ruth~S. Van~de Water.
\newblock Lattice qcd inputs to the ckm unitarity triangle analysis.
\newblock {\em Phys. Rev. D}, 81:034503, Feb 2010.

\bibitem{PhysRevD.84.014503}
Y.~Aoki, R.~Arthur, T.~Blum, et~al.
\newblock Continuum limit of ${B}_{K}$ from $2+1$ flavor domain wall qcd.
\newblock {\em Phys. Rev. D}, 84:014503, Jul 2011.

\bibitem{PhysRevD.87.094514}
R.~Arthur, T.~Blum, P.~A. Boyle, et~al.
\newblock Domain wall qcd with near-physical pions.
\newblock {\em Phys. Rev. D}, 87:094514, May 2013.

\bibitem{Blum:2012uh}
Thomas Blum, Taku Izubuchi, and Eigo Shintani.
\newblock {A new class of variance reduction techniques using lattice
  symmetries}.
\newblock 2012, arXiv:1208.4349.

\bibitem{Becirevic:2003wk}
Damir Becirevic and Giovanni Villadoro.
\newblock {Impact of the finite volume effects on the chiral behavior of f(K)
  and B(K)}.
\newblock {\em Phys.Rev.}, D69:054010, 2004, hep-lat/0311028.

\end{thebibliography}

\end{document}